\documentclass[11pt,twoside]{article}
\usepackage{newpasp,epsf,epsfig}
\def\edcomment#1{\iffalse\marginpar{\raggedright\sl#1\/}\else\relax\fi}
\reversemarginpar

\newcommand{\lya}{Ly{\sc $\alpha$}}
\newcommand{\ha}{H{\sc $\alpha$}}
\newcommand{\hb}{H{\sc $\beta$}}

\newcommand{\civ}{C\,{\sc iv}}

\newcommand{\ciii}{C\,{\sc iii}]}

\newcommand{\mgii}{Mg\,{\sc ii}}

\newcommand{\oiii}{[O\,{\sc iii}]}

\newcommand{\feii}{Fe\,{\sc ii}}

\begin{document}
\title{Principal Component Analysis of AGN Spectra}
\author{Zhaohui Shang}
\affil{University of Texas at Austin, Austin, TX 78712
%\\University of Wyoming, Laramie, WY 82070
}
\author{Beverley J. Wills}
\affil{University of Texas at Austin, Austin, TX 78712}

\begin{abstract}
We discuss spectral principal component analysis (SPCA) and show
examples of its application in analyzing AGN spectra in both small
and large samples.  It can be used to identify peculiar
spectra and classify AGN spectra.  Its application to correlation
studies of AGN spectral properties and spectral measurements for
large samples is  promising.

%We discuss spectral principal component analysis (SPCA) and show
%examples of its application in analyzing AGN spectra in both small
%and large samples.  It can be used to select peculiar spectra,
%classify spectra, measure spectra parameters and study correlations among
%spectral properties.  SPCA is especially a powerful tool in analyzing spectra
%of large samples.
%We demonstrate that SPCA is efficient in analyzing spectra of
%both small and large samples.  It can be used to identify peculiar
%spectra and classify AGN spectra.  Its application to correlation
%studies of AGN spectral properties and spectral measurements for
%large samples is  promising, but more investigations are needed.

\end{abstract}

\section{Introduction}

To study AGN spectra, one has to deal with many emission features as
well as continuum.  Simple statistics often cannot handle
the large number of measured parameters efficiently, therefore,
a multivariate analysis is needed.  Principal component analysis
(PCA) is one such powerful tool (see also Boroson 2004).

Suppose we have $n$ AGNs in a sample, each has $p$ measured parameters 
$X_1, X_2, \cdots, X_p$. We can write
{\small 
\[
\begin{array}{ccccc}
        & X_1   & X_2   & \cdots        & X_p   \\
QSO1    & x_{11}& x_{12}& \cdots        & x_{1p}\\
QSO2    & x_{21}& x_{22}& \cdots        & x_{2p}\\
\vdots  &      \vdots & \vdots &        & \vdots \\
QSOn    & x_{n1}& x_{n2}& \cdots        & x_{np}\\
\end{array}.
\]
}
\noindent
$X_1, X_2, \cdots, X_p$ are unit vectors and $x_{ij}$ are data.

PCA defines a set of new orthogonal variables, principal
components $W_j$ ($j=1, \cdots, p$), which are linear combinations 
of the original variables\\
%\begin{equation}
\makebox[5in][c]{
$W_j = e_{j1}X_1 + e_{j2}X_2 + \cdots + e_{jp}X_p ,$ 
}
%\end{equation}
\hfill (1)
\\
An easy way to understand PCA is to consider it from the geometrical
point of view (Francis \& Wills 1999).  
PCA aims to define orthogonal principal axes in a
multi-dimensional space, as shown in Fig.~1, where a correlation
exists between measured parameters $X_1$ and $X_2$, and two principal
components (PCs) are defined.  PC1 (or $W_1$) accounts for most variance (the
correlation), PC2 (or $W_2$) is small and can be ignored.
\begin{figure}
\plotfiddle{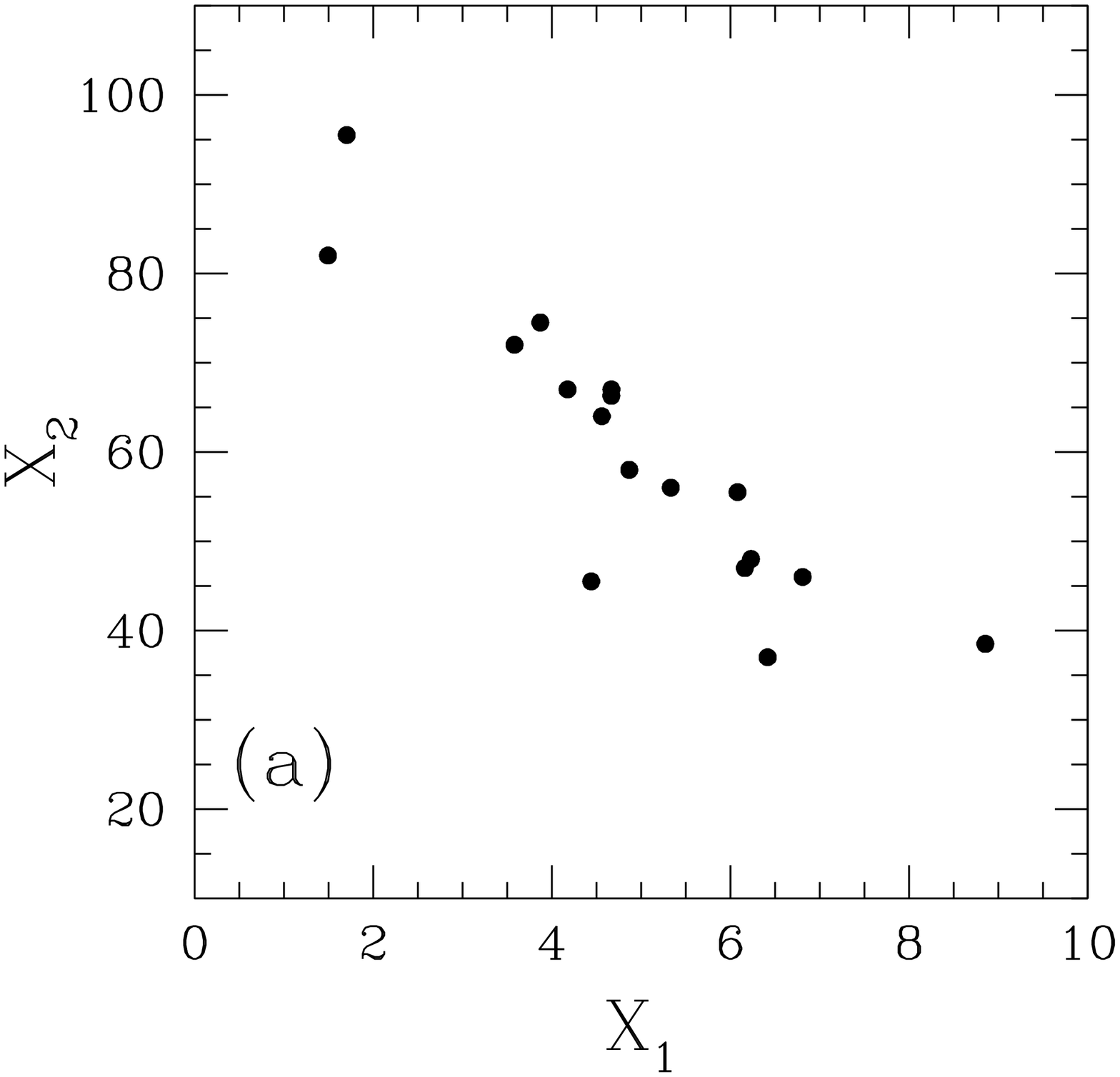}{.8in}{0}{20}{20}{-142}{-64}
\plotfiddle{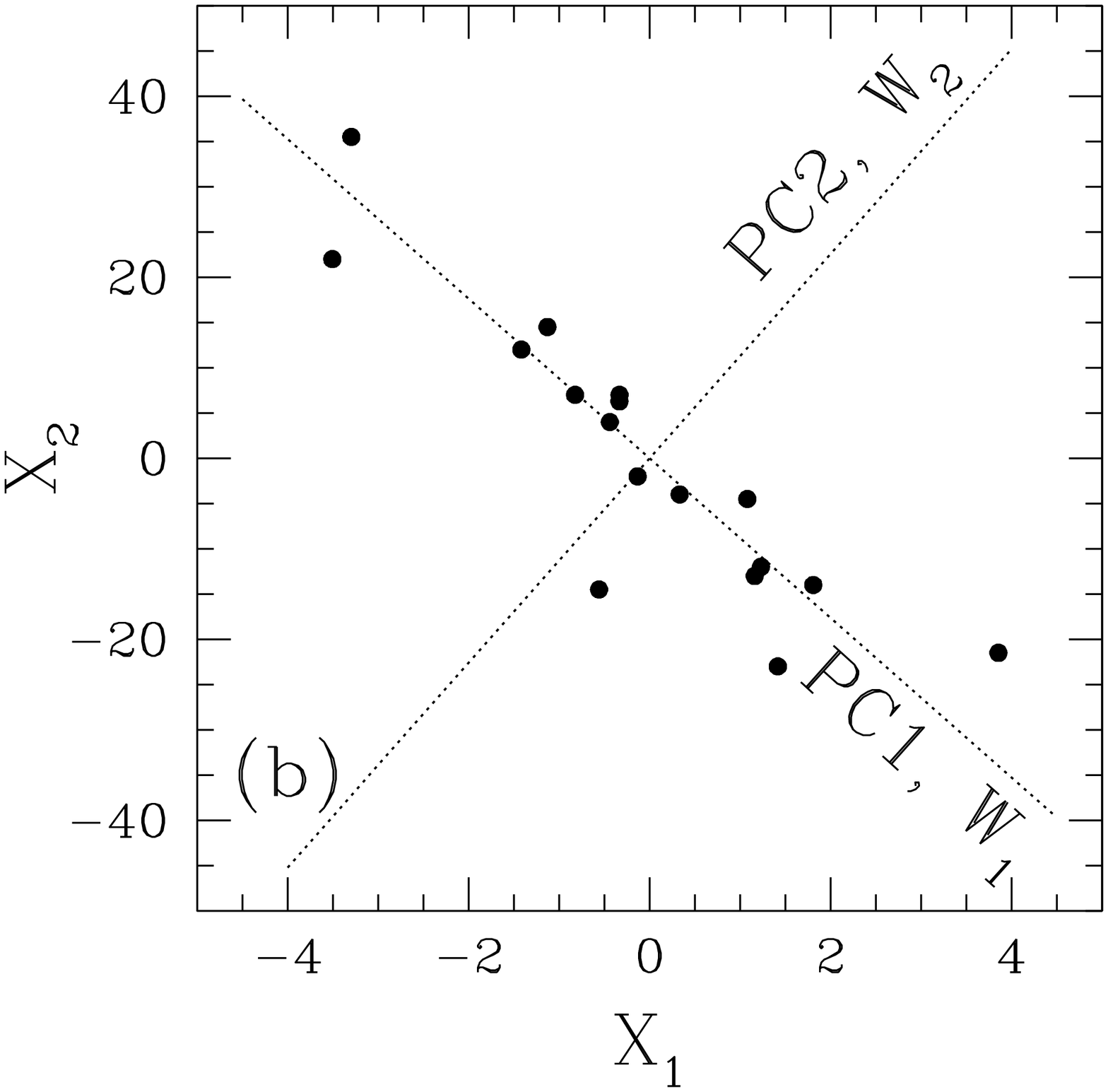}{.0in}{0}{20}{20}{0}{-40}
\caption
{Geometrical point of view of PCA.  (a) Measured
values.  (b) Mean subtracted values.  Two principal axes PC1 and PC2 are
defined.}
\end{figure}

%\begin{figure}[t]
%\begin{minipage}[t]{0.5\textwidth}
%    \begin{center}
%    \epsfig{file=pcageoa.eps, scale=0.15}
%    \end{center}
%\end{minipage}
%\hfill
%\begin{minipage}[t]{0.5\textwidth}
%    \begin{center}
%        \epsfig{file=pcageob.eps, scale=0.15}
%    \end{center}
%\end{minipage}
%\caption[Geometrical Point of View of PCA.]
%{Geometrical point of view of PCA.  (a) Measured
%values.  (b) Mean subtracted values.  Two principal axes PC1 and PC2 are
%defined.}
%%PC1 is significant, and PC2 can be ignored.}
%\label{pcageo}
%\end{figure}

PCA can be applied to AGN spectra directly (Francis et al. 1992; Shang
et al. 2003).  In this spectral principal component analysis (SPCA),
each spectrum is divided into small wavelength bins, and the flux in
each bin (e.g., $F_{\lambda_i}$) is an input variable.  Eq.~(1)
can be written as\\
%\begin{equation}
\makebox[5in][c]{
$W_j = e_{j1}F_{\lambda_1} + e_{j2}F_{\lambda_2} + \cdots +
e_{jp}F_{\lambda_p} ,$
}
%\end{equation}
\hfill (2)
\\
The result principal components can then be also
represented as spectra,
$SPC_j = [e_{j1}\  e_{j2}\   \cdots \  e_{jp} ]$,
namely, the spectral principal components (SPCs). 
%\begin{equation}
%\label{eqSPCj}
%SPC_j = [e_{j1}\  e_{j2}\   \cdots \  e_{jp} ] ,
%\end{equation}
An original spectrum can be reconstructed by adding the
weighted principal component spectra to the mean spectrum,\\
%\begin{equation}
\makebox[5in][c]{
$
Spectrum_i = Mean + w_{i1} {SPC_1} + w_{i2} {SPC_2} + \cdots + w_{ip} {SPC_p},
$
}
%\end{equation}
\\
where $w_{ij}$ are calculated from Eq.~2 for each object $i$,
and are referred as the weights (or scores) of $SPC_j$.
In other words, all original spectra are made from the
principal component spectra.  
This implies that a limited number of SPCs can be
used to measure spectra in large samples (see also Yip et al. 2004).
%This implies that spectra can be measured from a limited number
%of significant SPCs (see also Yip et al. 2004).

If there are strong linear relationships among the
original variables (or wavelength bins), 
each of these relationships will be represented by a
principal component.  Fewer principal components may be required to
describe the total variation of a sample, thus providing a simpler
description of the dataset (Francis et al. 1992).  Any principal
component accounting for a significant fraction of the total sample
variance might be related to one or more underlying fundamental physical
parameters, giving some physical insight into the cause of the
variations (Boroson \& Green 1992, Boroson 2002).
However, PCA is only a statistical tool and 
it is still up to the investigators to judge whether the resulting
PCs have any physical meanings.

PCA is a linear analysis.  If there are strong non-linear
relationships involved, PCA is not able to identify them directly and there
will be crosstalk among the resulting principal components.  When a
non-linear part of a relationship is not strong, PCA is still able to 
follow the ``linear'' trend of the relationship.  In SPCA, good redshifts
are needed because line shifts can cause strong non-linear effects.
Line width change also introduces non-linear relationships among binned
fluxes, so caution is needed in interpreting SPCA results.

%\section{Spectral Principal Component Analysis}

\section{SPCA on a Small Sample of Quasars}

Shang et al.\ (2003) show the power of SPCA in analyzing a
small sample of QSOs.  SPCA decomposes the QSO spectra into
three independent, significant principal components (Fig.~2).
SPC1 represents the Baldwin effect, but only line-cores are involved.
By using SPC1 instead of integrated line EW, the scatter in the 
luminosity relationship is reduced.  SPC2 shows
changes of UV-optical continuum slope, due to intrinsic QSO continuum
slope variation and/or reddening.  SPC3 extends Boroson \& Green's
Eigenvector 1 relationship (Boroson \& Green 1992), to include many
UV line properties and line-width changes.

\begin{figure}[t]
\plotfiddle{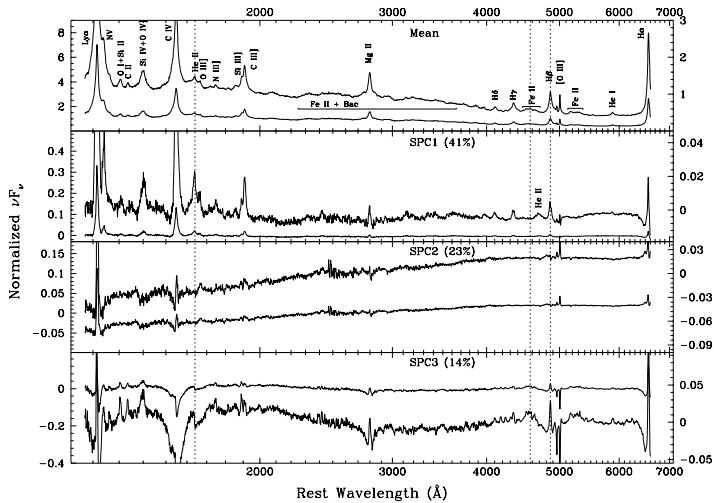}{2.3in}{270}{36}{36}{-144}{190}
\caption
{Mean spectrum and the three significant spectral principal components
from SPCA for the small sample (Shang et al. 2003). The numbers in parenthesis are the
fractions of the sample variance accounted for by each principal
component.  Spectral features in opposite directions in the principal
components are anti-correlated, and vice versa. The ``W'' shape of \lya,
\mgii, \hb, and \ha\ in SPC3 indicates that the width increases with stronger
\oiii. 
The very broad ``small blue bump'' is anticorrelated with
the optical \feii\ features.
%The very broad ``small blue bump'' appears in SPC3. 
}
\end{figure}

%\begin{figure}[t]
%\begin{center}
%\epsfig{file=allpcland.eps,height=3.8in,angle=270}
%\caption
%{Mean spectrum and the three significant spectral principal components
%from SPCA for the small sample (Shang et al. 2003). The numbers in parenthesis are the
%fractions of the sample variance accounted for by each principal
%component.  Spectral features in opposite directions in the principal
%components are anti-correlated, and vice versa. The ``W'' shape of \lya,
%\mgii, \hb, and \ha\ in SPC3 indicates that the width increases with stronger
%\oiii. 
%The very broad ``small blue bump'' is anticorrelated with
%the optical \feii\ features.
%%The very broad ``small blue bump'' appears in SPC3. 
%}
%\end{center}
%\end{figure}

\section{SPCA on Large Samples}

One of the advantages of SPCA is that correlations can be
investigated without parameterizing the line profiles or defining
the continua.  Therefore, it is especially useful for analyzing
large samples.

Also unlike the ``composite spectrum analysis'', SPCA analysis
keeps the information for individual objects, e.g., the weights
of the principal component spectra for each object, which can be
used to correlate with other non-spectral properties, such as black
hole mass etc.  SPCA has also been used for classifying AGN spectra
(e.g., Francis et al. 1992, Boroson 2002).

To demonstrate its use for large samples, we applied SPCA to the
SDSS DR1 QSO spectra in the region of \lya-\civ-\ciii.  We choose
only 771 objects with high redshift-confidence ($\verb"z_conf" >
0.95$) (Stoughton et al. 2002).  For a sample with uniform spectral
properties, the distribution of the weights of SPCs should be random
and centered at the origin.  However, there are outliers in the distribution
of the weights of SPC1 and SPC2 (Fig.~3, left), indicating that these
spectra are peculiar (Fig.~3, right).
\begin{figure}[t]
\plotfiddle{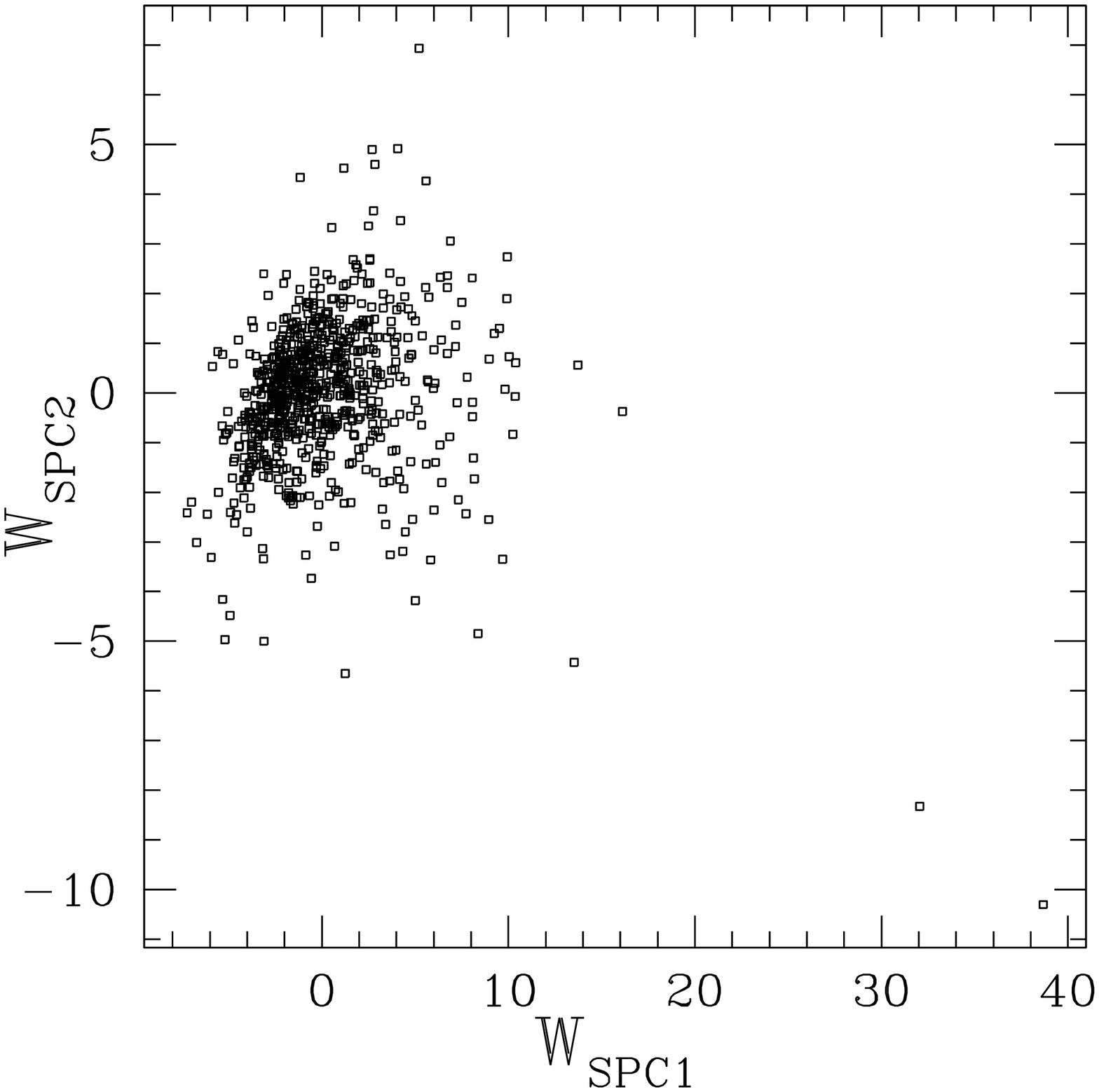}{.87in}{0}{20}{20}{-152}{-64}
\plotfiddle{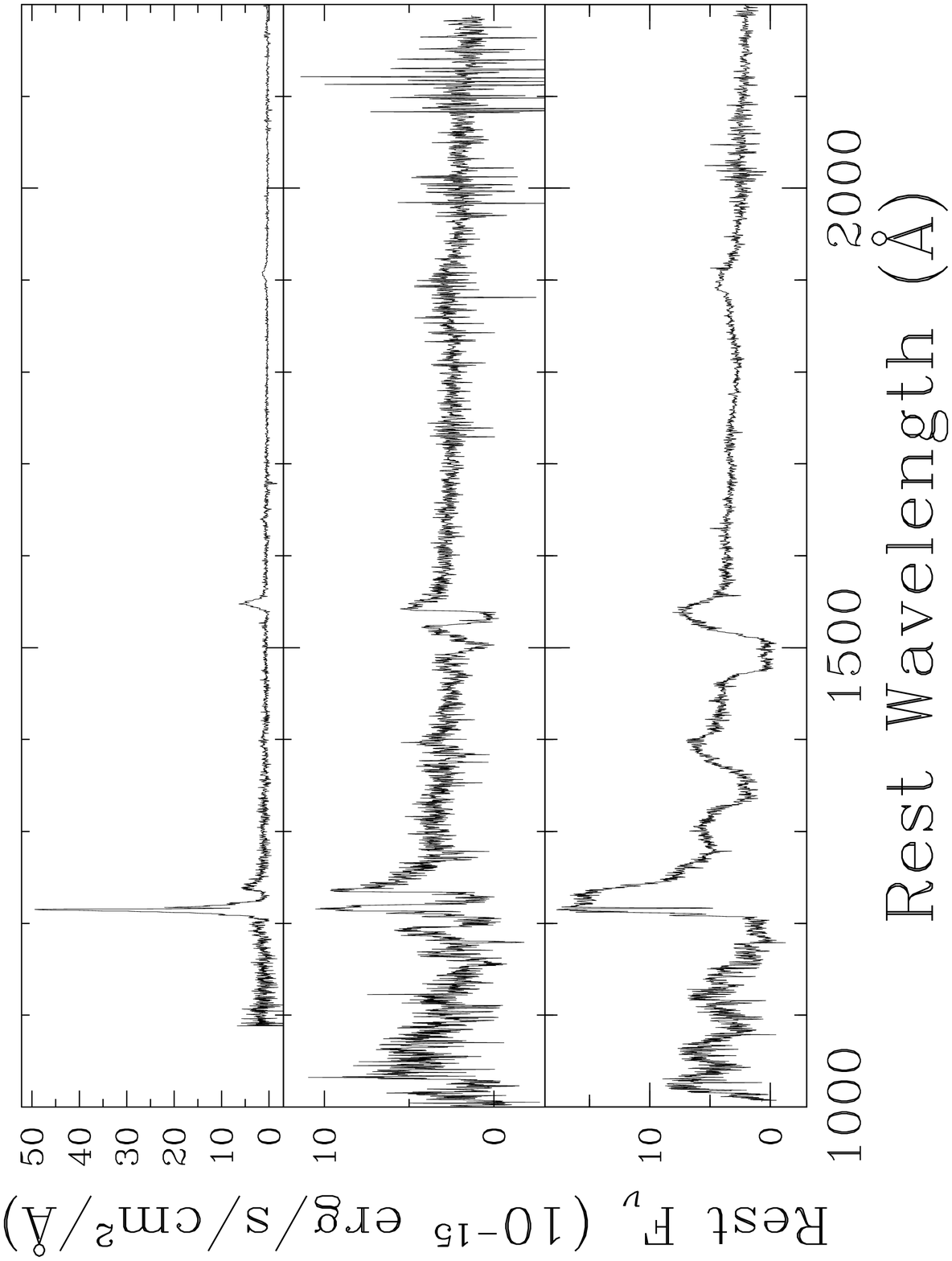}{.0in}{270}{20}{20}{0}{110}
\caption{Left: Distribution of the weights of SPC1 and SPC2 with all 771
spectra.  The outliers have peculiar spectra.  Right: Examples of
peculiar spectra selected with SPCA.}
\end{figure}

%\begin{figure}[t]
%\begin{minipage}[t]{0.50\textwidth}
%    \begin{center}
%    \epsfig{file=Hza03Weights.eps, scale=0.15}
%    \end{center}
%\end{minipage}
%\hfill
%\begin{minipage}[t]{0.50\textwidth}
%    \begin{center}
%        \epsfig{file=Hzspec.eps, scale=0.15}
%%        \epsfig{file=Hzspec.eps, height=1.00in, angle=270}
%    \end{center}
%\end{minipage}
%\caption{Left: Distribution of the weights of SPC1 and SPC2 with all 771
%spectra.  The outliers have peculiar spectra.  Right: Examples of
%peculiar spectra selected with SPCA.}
%\end{figure}

We are interested in the spectral properties of the majority of the
sample, so after excluding the peculiar spectra, SPCA is applied
to the remaining 639 spectra.  Fig.~4 (left) shows the first three
principal components which can be used to classify the spectra and
investigate the correlations among the spectra properties.  
SPC1 shows
that the line-cores are correlated with each other, similar to the
SPC1 in Sec. 2, but the expected Baldwin effect has large scatter
(Fig.~4, right).  SPC2 shows line-width change: strong-SPC2 objects
have narrow \lya.  SPC3 shows absorptions (or line shifts) in \lya\
and \civ; strong-SPC3 objects have \lya\ and \civ\ absorptions.

\begin{figure}
\plotfiddle{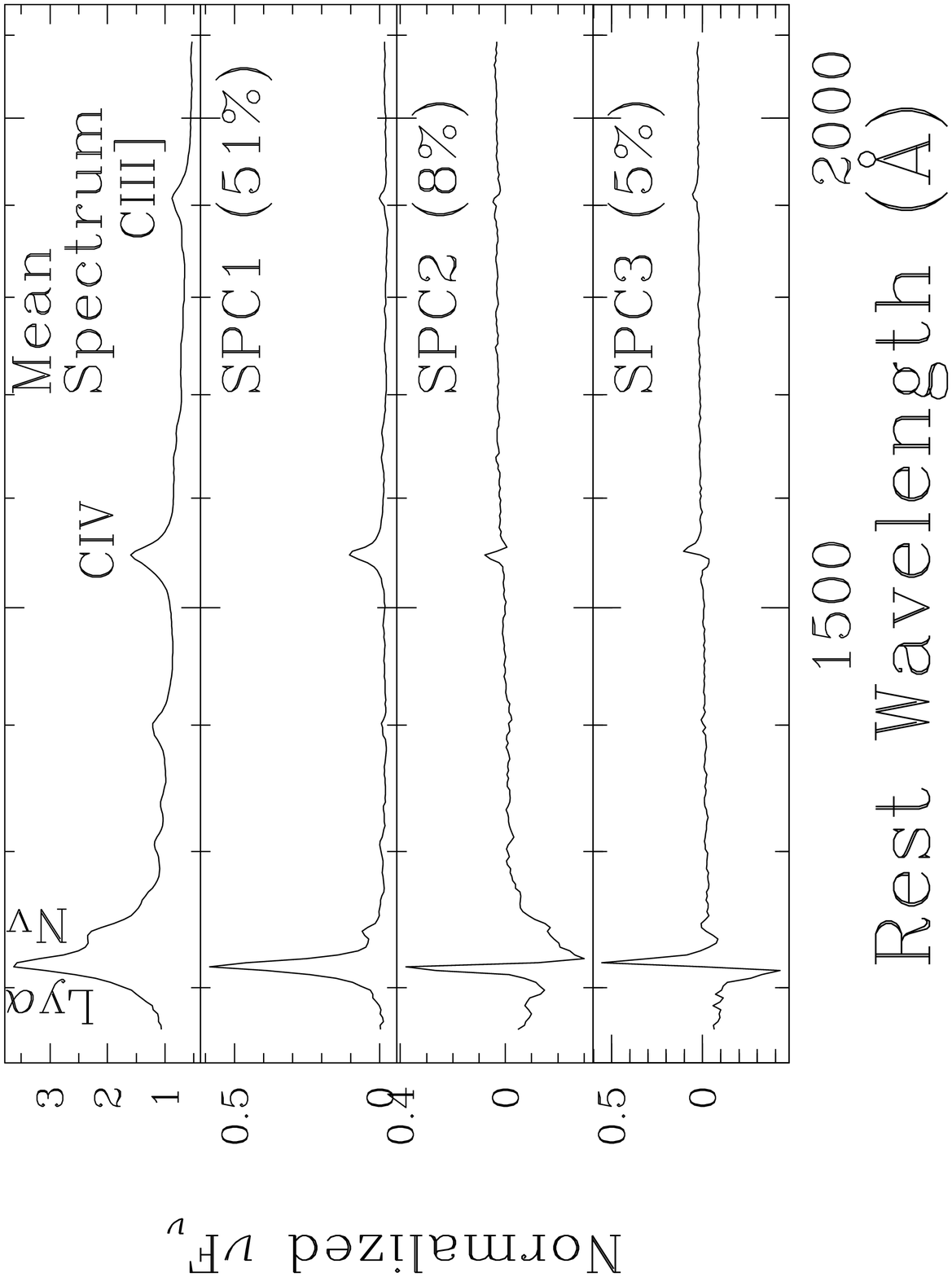}{.90in}{270}{20}{20}{-162}{84}
\plotfiddle{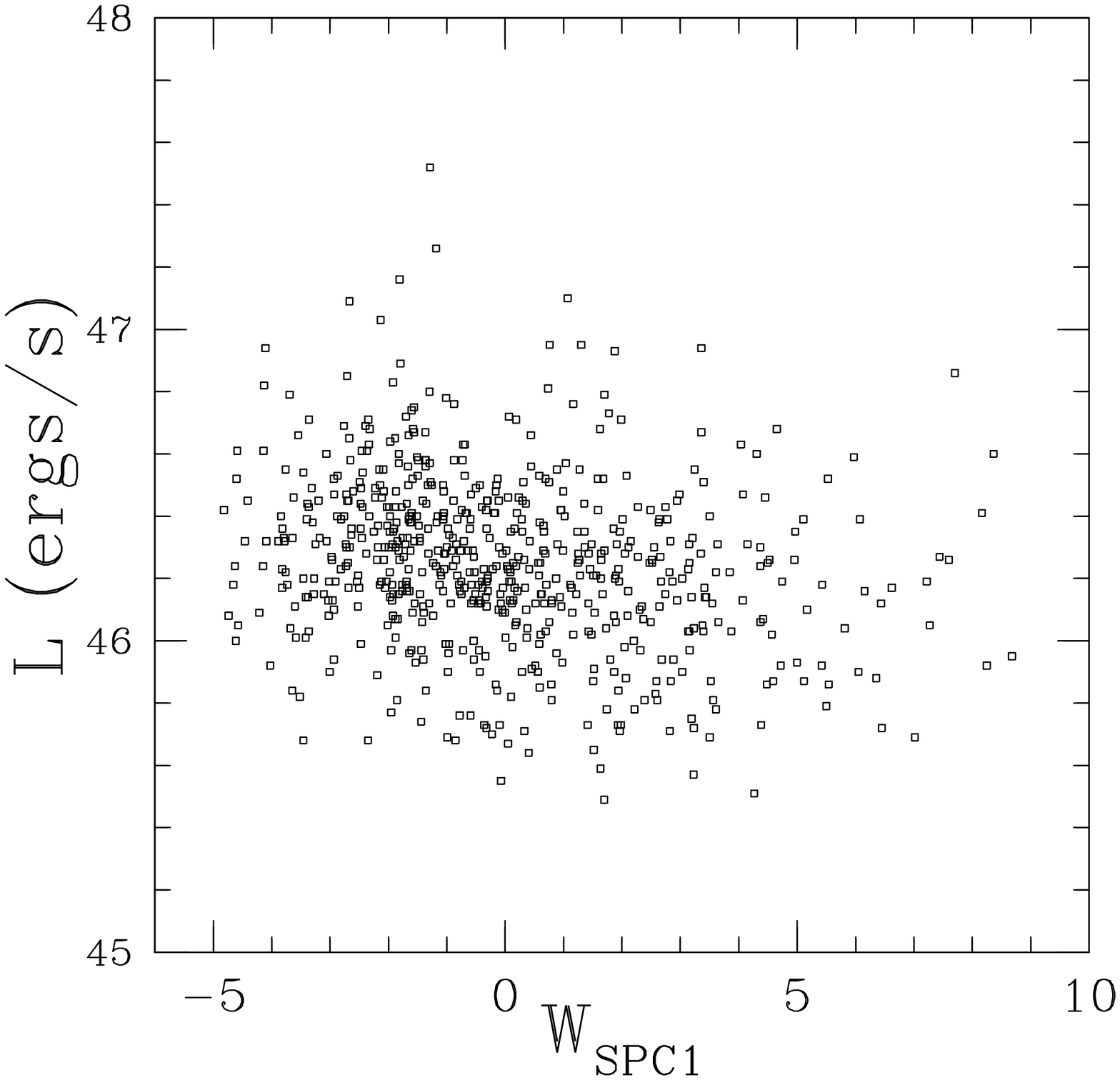}{.0in}{0}{20}{20}{35}{-40}
\caption{Left: SPCA results for 639 spectra.  Right: Continuum
luminosity vs.\ SPC1 (line-core).  The scatter in the anticorrelation
($P<10^{-4}$) is large.  The luminosity is calculated with the measured
continuum flux at 1682\AA\ assuming H$_0$=50, q$_0$=0.5.}
\end{figure}

%\begin{figure}
%\begin{minipage}[t]{0.50\textwidth}
%    \begin{center}
%    \epsfig{file=defHzpca.eps, scale=0.15}
%    \end{center}
%\end{minipage}
%\hfill
%\begin{minipage}[t]{0.50\textwidth}
%    \begin{center}
%        \epsfig{file=HzBeff.eps, scale=0.15}
%    \end{center}
%\end{minipage}
%\caption{Left: SPCA results for 639 spectra.  Right: Continuum
%luminosity vs.\ SPC1 (line-core).  The scatter in the anticorrelation
%($P<10^{-4}$) is large.  The luminosity is calculated with the measured
%continuum flux at 1682\AA\ assuming H$_0$=50~km/s/Mpc, q$_0$=0.5.}
%\end{figure}

%\section{Conclusions} 

%We demonstrate that SPCA is efficient in analyzing spectra of
%both small and large samples.  It can be used to identify peculiar
%spectra and classify AGN spectra.  Its application to correlation
%studies of AGN spectral properties and spectral measurements for
%large samples is  promising, but more investigations are needed.
%
%Future work will include detailed SPCA analyses of different type
%spectra, in different redshift and luminosity bins etc.

%\section{References}

\end{document}